\documentclass[a4paper]{proc}
\usepackage{graphicx}

\makeatletter
\newcommand{\textsubscript}[1]{%
	\ensuremath{_{\mbox{\fontsize{\sf@size}{0}\selectfont#1}}}}
\@ifundefined{textsuperscript}{\newcommand{\textsuperscript}[1]{%
	\ensuremath{^{\mbox{\fontsize{\sf@size}{0}\selectfont#1}}}}}{}
\makeatother

\begin{document}


\title{Excitations in Spin Chains and Specific-Heat Anomalies in
	Yb\textsubscript4As\textsubscript3}

\author{Burkhard Schmidt\thanks{E-mail: {\tt bs@mpipks-dresden.mpg.de}},
	Peter Thalmeier and Peter Fulde\\
	{\em Max-Planck-Institut f\"ur Physik komplexer Systeme,}\\
	{\em D-01187 Dresden, Germany}}

\maketitle

\begin{abstract}
An explanation is given for the observed magnetic-field dependence of
the low-temperature specific heat coefficient of
Yb\textsubscript4As\textsubscript3. It is based on a recently
developed model for that material which can explain the observed
heavy-fermion behaviour. According to it the
Yb\textsuperscript{3+}-ions are positioned in a net of parallel chains
with an effective spin coupling of the order of $J\approx25\,\rm
K$. The magnetic-field dependence can be understood by including a
weak magnetic coupling $J'$ between adjacent chains. The data require
a ratio $J'/J\approx10^{-4}$. In that case the experimental results
can be reproduced very well by the theory.
\end{abstract}

\section*{ }
The low-carrier compound Yb\textsubscript4As\textsubscript3 which
belongs to the Rare-Earth pnictide family has been subject of
extensive experimental investigation caused by its intriguing
low-temperature
anomalies~\cite{bonville94,helfrich95,ochiai90,goto95,mori95,krolas95,kohgi95,suga95,reinders93}.
It shows evidence for a `heavy-fermion' type behaviour below a
characteristic temperature $T^\ast\simeq40\,\rm K$, notably a large
$\gamma$-coefficient of about $200\,\rm mJ/mol\cdot K^2$ for the
specific heat. On the other hand the carrier concentration is
extremely small, approximately $10^{-3}$ holes per formula unit below
$T^\ast$. This excludes a conventional Kondo-mechanism as origin of
the low-energy excitations leading to the large $\gamma$-value.

The scenario proposed in Ref.~\cite{fulde95} is quite different from
that of a Kondo system. The crystal has the
anti-Th\textsubscript3P\textsubscript4-structure, and the space group
is I\=43d. The Yb-ions occupy four systems of parallel chains oriented
along the space diagonals of a cube (body-centred cubic rod
packing)~\cite{okeeffe77}. A structural phase transition at room
temperature changes the high-temperature homogeneous mixed-valent
state of Yb\textsubscript4As\textsubscript3
($\mbox{Yb\textsuperscript{3+}}:\mbox{Yb\textsuperscript{2+}}=1:3$) to
a charge ordered state where the $25\,\%$ Yb\textsuperscript{3+}-holes
are confined to one of the four chain systems, e.g, parallel
$\langle111\rangle$-chains. The particular ordering of
Yb\textsuperscript{3+}-ions was confirmed clearly by perturbed
angular-correlation (PAC) measurements~\cite{krolas95}. This state has
almost half-filled hole bands which are subject to strong on-site
Coulomb correlations, effectively described by a t-J model. Therefore
the low-energy excitations essential for the specific heat are the
magnetic excitations of an antiferromagnetic quasi-one-dimensional
`$S=\frac{1}{2}$' Heisenberg chain ($\mathbf S$ is the pseudospin
corresponding to the lowest Kramers doublet of
Yb\textsuperscript{3+}). This model proposed in Ref.~\cite{fulde95}
was beautifully confirmed by the inelastic neutron-scattering results
of Kohgi et al.~\cite{kohgi95}. They show that low-temperature
magnetic excitations in Yb\textsubscript4As\textsubscript3 are very
well described by the one-dimensional spin wave spectrum,
$\omega(q)=(\pi/2)J\sin(aq)$, of des Cloizeaux and
Pearson~\cite{cloizeaux62} where q is the wave number in
$\langle111\rangle$-direction and $a$ the distance of
Yb\textsuperscript{3+} ions along the chain. This also leads naturally
to a linear specific heat for $k_{\rm B}T/J\ll1$. From the maximum
observed spin-wave energy of $3.8\,\rm meV\approx40\,\rm K$ at
$q=\frac{1}{2}(\pi/d)$ one obtains $J=25\,\rm K$ in good agreement
with the energy scale $T^\ast$ estimated from the value of~$\gamma$.

However there is still one major challenge to the validity of the
model proposed in Ref.~\cite{fulde95}. Concerning the dependence of
$\gamma$ on an applied magnetic field $\mathbf H$, one would expect
deviations from $\gamma(\mathbf H=0)$ of the order $(\mathbf H/J)^2$,
i.e., about $10^{-2}$ at $4\,\rm T$ uniformly for all temperatures
with $k_{\rm B}T/J\leq1$. Experiments by Helfrich et
al.~\cite{helfrich95} revealed a quite different behaviour: Above
$2\,\rm K$ changes in $\gamma$ are indeed small. In contrast,
at $0.5\,\rm K$ the field dependence of $\gamma(\mathbf H)$ is
dramatic. Already at $4\,\rm T$ its value is strongly suppressed in
conflict with the above argument. This indicates that an additional
energy scale much smaller than $J$ ($25\,\rm K$) must be present in
Yb\textsubscript4As\textsubscript3.

It is the purpose of this letter to show that the spin-chain model for
Yb\textsubscript4As\textsubscript3 is well capable of explaining the
unexpected results of Helfrich et al. For the present purpose we may
neglect deviations from perfect charge ordering or half filling which
is only important for the transport properties. The additional energy
scale which the results in Ref.~\cite{helfrich95} suggest is
associated with anisotropy gaps of the magnetic excitations caused by
interchain coupling $J'$ of the parallel Yb\textsuperscript{3+}-spin
chains. The complete excitation spectrum of a trigonal
antiferromagnetically coupled chain system is calculated and it is
shown how the linear specific-heat term and its field dependence
arises. A comparison with the data of Helfrich et al.\ allows to
estimate the magnitude of $J'$. Proposals for magnetic resonance
experiments are made.

It is well-known that in one spatial dimension, there cannot be any
long-range spin ordering, but, according to calculations on the
spin-1/2 Heisenberg chain~\cite{affleck89,hallberg95}, the spin-spin
correlation function decays slowly, $\langle\mathbf S(r)\cdot\mathbf
S(0)\rangle\propto(-1)^r\sqrt{\log(r)}/r$, i.e., the correlation
length is infinite. We therefore model
Yb\textsubscript4As\textsubscript3 as if the spins were
antiferromagnetically ordered and apply linear spin-wave theory,
neglecting disorder-induced corrections.

We transform the crystal coordinate system of the charge-ordered state
to a new basis with a hexagonal unit cell where the
$\langle111\rangle$-direction now is oriented along the c-axis, i.e.,
it transforms to the $\langle001\rangle$-direction.  Taking into
account the interchain coupling by a parameter $\Delta$, we describe
Yb\textsubscript4As\textsubscript3 with a Hamiltonian of the form
\begin{eqnarray}
	H&=&\sum_{\langle ij\rangle}\mathbf S_i\cdot\mathbf S_j
		+\Delta\sum_{\langle ij\rangle'}\mathbf
		S_i\cdot\mathbf S_j
\nonumber\\
	&+&\sum_i\left(d\left(S_i^z\right)^2 
			-\mathbf h\cdot\mathbf S_i\right),
\label{eqn:hamiltonian}
\end{eqnarray}
where $\Delta=J'/J$ is the ratio of the antiferromagnetic exchange
constants perpendicular and parallel to the c-axis, and $\mathbf
h=g_\textrm{\scriptsize eff}\mu_{\mathrm B}\mathbf H/J$ is the applied
magnetic field. All energies are in units of $J$. We parameterize the
anisotropic exchange of the system by introducing a parameter $d>0$
describing a single-ion easy-plane anisotropy. The symbols $\langle
ij\rangle$ and $\langle ij\rangle'$ imply summation over
nearest-neighbour bonds along and perpendicular to the c-axis.

Starting point of the calculation is the classical
N\'eel-state. Perpendicular to the c-direction, the spins are situated
on a trigonal lattice-structure, subdivided into three
interpenetrating sublattices. These sublattices are stacked
antiferromagnetically along the c-direction. The ordered system thus
contains six sublattices $\mathcal L(L,c)$, numbered with $L=1,2,3$
within one plane perpendicular to the c-axis, and with $c=1,2$ in
c-direction. In order to describe the N\'eel-state properly, we
perform coordinate transformations on each sublattice $\mathcal
L(L,c)$ such that each $\mathbf S_i$ points into the local
$z$-direction. Parametrizing these transformations by polar and
azimuthal angles $\Phi_{L,c}$ and $\Theta_{L,c}$, we apply a
Holstein-Primakoff transformation on the spins and retain only the
constant and bilinear parts of the Hamiltonian. After Fourier
transforming $H$, we diagonalize it by a
Bogoliubov-transformation. The diagonal form of $H$ then reads
\begin{eqnarray}
	H&=&H\textsubscript{N\'eel}+H_0
\nonumber\\
	&+&2S\sum_{L,c}\sum_{\mathbf q} \omega_{Lc}(\mathbf q)
			\left(a_{q,Lc}^\dagger a_{q,Lc}+\frac{1}{2} \right).
\label{eqn:h_bilinear}
\end{eqnarray}
$H_0\propto(S,h)$ contains constants arising from a rearrangement of
the bilinear terms of $H$, and $a_{q,Lc}$ ($a_{q,Lc}^\dagger$) is the
Fourier transform of the Holstein-Primakoff boson $a_i$
($a_i^\dagger$) on the sublattice $\mathcal
L(L,c)$. $H\textsubscript{N\'eel}\propto(S^2,h\cdot S)$ contains the
classical part of $H$, i.e, the exact Hamiltonian for
$S\to\infty$. Minimizing it with respect to the angles $\Phi_{Lc}$ and
$\Theta_{Lc}$ yields the ground-state spin configuration of
(\ref{eqn:h_bilinear})~\cite{chubukov90,rastelli94}. The problem of
quantum corrections to the ground-state energy may be important for
some cases~\cite{shiba93,shiba94}.

If the applied field $\mathbf h$ does not break the rotational
symmetry of (\ref{eqn:h_bilinear}) in the limit $\mathbf h=0$, i.e, for
$\mathbf h$ parallel to the c-axis, the N\'eel-state is made up of two
antiferromagnetically stacked triangular lattices. Within these
lattices, the spins of the three sublattices lie on the surface of a
cone ($|\Phi_{Lc}-\Phi_{L'c}|=2\pi/3$, $L\neq L'$) with the
axis parallel to the field (the c-axis) and an apex angle
$\Theta_{Lc}=\Theta=\arccos\left(\frac{h/(2S)}{2+d+9/2\Delta}\right)$.
The spin-wave dispersion is given by
\begin{eqnarray}
\omega_{Lc}^2(\mathbf q)&=&
	\left(1-2\cos^2\Theta+\frac{h}{2S}\cos\Theta\right.
\nonumber\\
	&\phantom{\times}&{}+(-1)^c\cos q_z
\nonumber\\
&\phantom{\times}&\left.{}+\frac{3}{2}\Delta
	\left(1-3\cos^2\Theta-2\rho_L(\mathbf q)\right)\right)
\nonumber\\
&\times&\left(1+d-2\cos^2\Theta+\frac{h}{2S}\cos\Theta\right.
\nonumber\\
	&\phantom{\times}&{}-(-1)^c\left(1-2\cos^2\Theta\right)\cos q_z
\nonumber\\
&\phantom{\times}&{}+\frac{3}{2}\Delta
	\left(1-3\cos^2\Theta\vphantom{\frac{3}{2}}\right.
\nonumber\\
&\phantom{\times}&\left.\left.{}
	+4\left(1-\frac{3}{2}\cos^2\Theta\right)
		\rho_L(\mathbf q)\right)\right),
\\[\baselineskip]
\rho_L(\mathbf q)&=&\frac{1}{3}\left(
		\cos^2\left(\frac{q_x}{2}-\Phi_L\right)
		\vphantom{\cos\left(\frac{\sqrt{3}q_y}{2}\right)}\right.
\nonumber\\
&\phantom{=}&{}\left.
		+\cos\left(\frac{q_x}{2}-\Phi_L\right)
			\cos\left(\frac{\sqrt{3}q_y}{2}\right)
		-\frac{1}{2}\right),
\nonumber\\
	& &\Phi_1=0,\quad\Phi_2=\frac{2\pi}{3},\quad
		\Phi_3=-\frac{2\pi}{3}.
\nonumber
\end{eqnarray}
According to Ref.~\cite{rastelli94}, with a magnetic field
perpendicular to the c-axis, one obtains for a small single-ion
anisotropy $d$ a so-called umbrella-type spin configuration for
$h>h_1\approx2S\sqrt{d}$. In the umbrella-phase, the spins cant out of
the c-plane---they lie on the surface of two cones with the axis
parallel to the field direction, but different apex angles. Since in
this case a simple closed-form solution for $\omega_{Lc}(\mathbf q)$
is not achievable, we diagonalized the
Hamiltonian~(\ref{eqn:h_bilinear}) numerically.

\begin{figure}
\vspace*{-\baselineskip}
\centerline{\includegraphics[width=7cm]{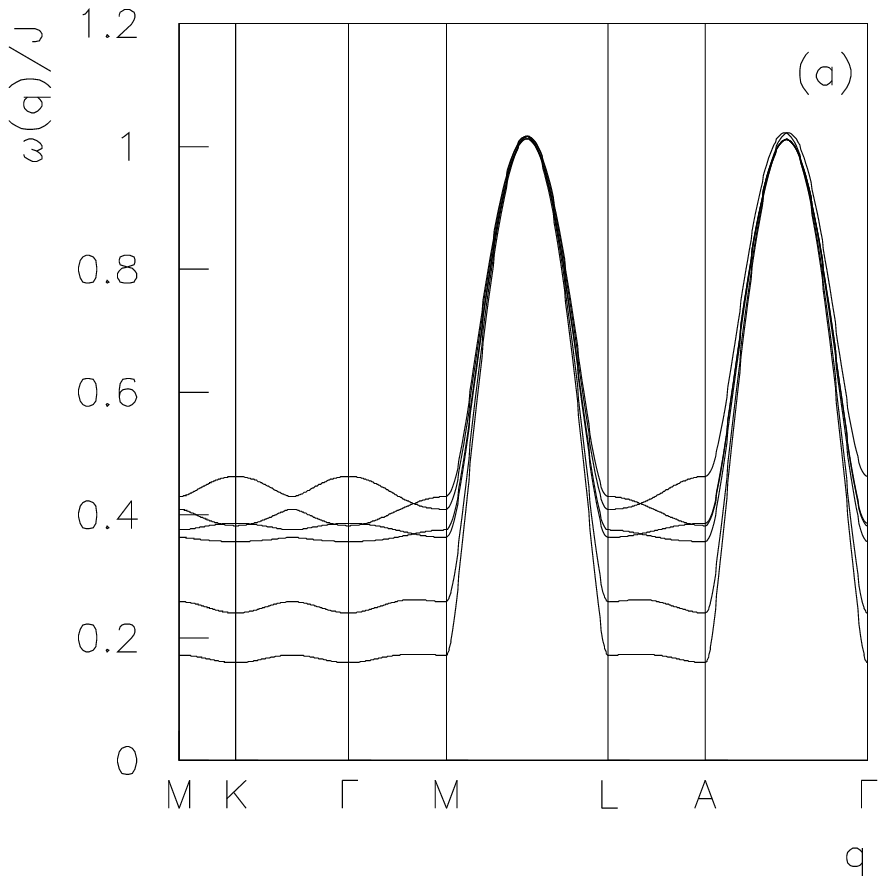}}
\vspace*{-0.5\baselineskip}
\centerline{\includegraphics[width=7cm]{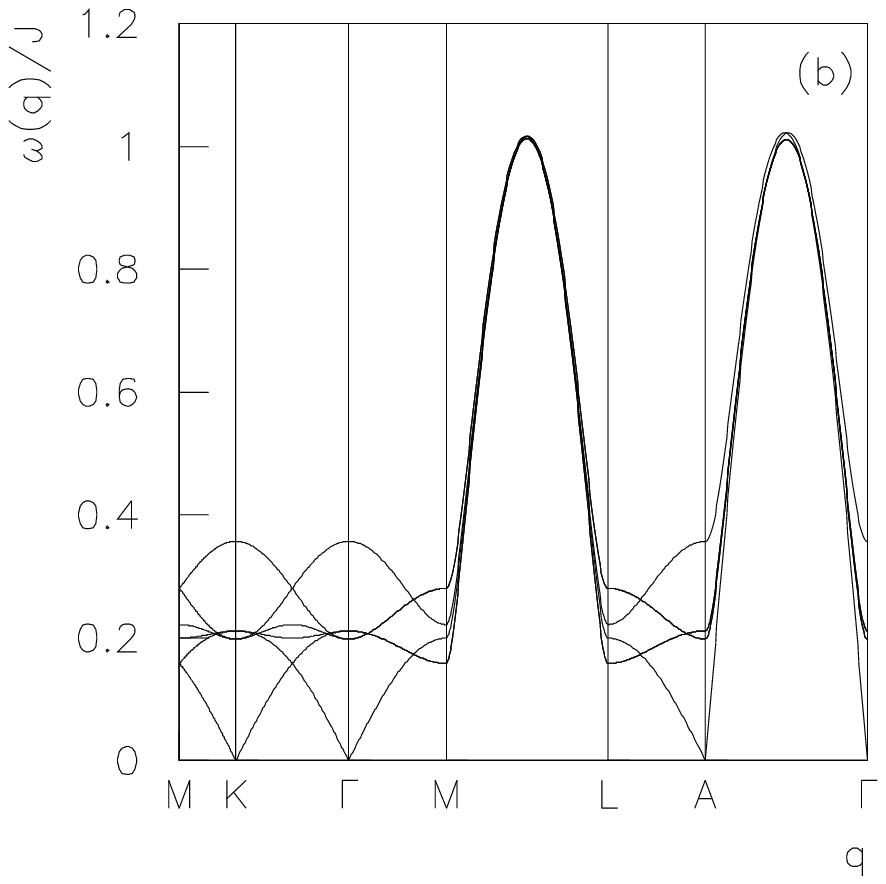}}
\vspace*{-0.5\baselineskip}
\centerline{\includegraphics[width=7cm]{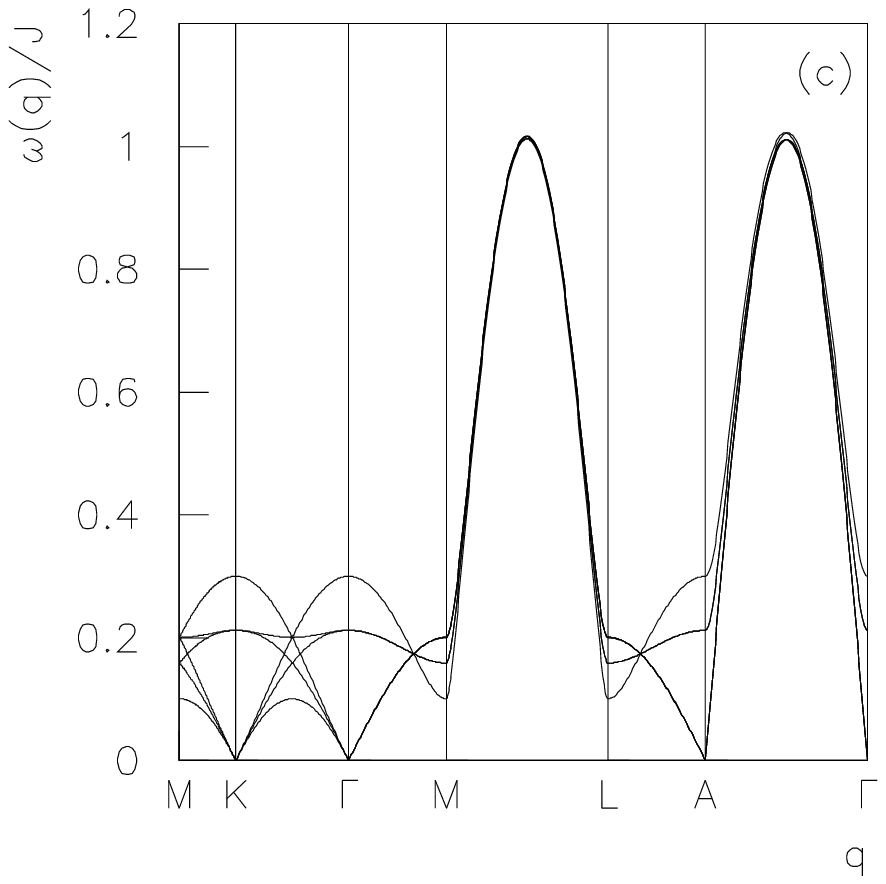}}
\caption{Spin-wave dispersion for the trigonal-chain structure in a
	magnetic field. (a) Field applied perpendicular to the c-axis,
	(b) field parallel to the c-axis, (c) without applied
	field. The ratio $\Delta$ of interchain-to intrachain-coupling
	is 0.01.}
\label{fig:sw}
\end{figure}
Figure~\ref{fig:sw}c shows a typical dispersion unfolded in the
chemical Brillouin zone for an interchain coupling $\Delta=0.01$,
vanishing easy-plane anisotropy $d$, and vanishing magnetic field. In
the directions parallel to the c-axis, the scale is set by the
intrachain coupling, which fits the neutron scattering data on
Yb\textsubscript4As\textsubscript3 if we set $J\approx25\,\mathrm K$,
rescaling by a factor of $2/\pi$ according to the Bethe-Ansatz
result. In the directions transverse to the c-axis, the scale of the
dispersion of the spin-wave modes is given by $\sqrt{\Delta}$. There
are three modes becoming soft at the $\Gamma$-point, corresponding to
the continuous degeneracy of the ground state with respect to
rotations of all spins by the same angle around an arbitrary axis.

Upon applying a magnetic field, this degeneracy is lifted, and the
corresponding modes acquire a finite frequency at the $\Gamma$-point.
For an arbitrary small but finite anisotropy $d$, we have still one
Goldstone mode if the field is applied parallel to the c-axis,
corresponding to the independence of the ground state with respect to
rotations around that axis, see Figure~\ref{fig:sw}b. For a field
direction perpendicular to the c-axis, a gap opens in the entire
Brillouin zone. For $\sqrt{d}\ll h\ll h_{\rm s}=2+9\Delta/2$, the
value of the gap is $E_{\rm g}/J\approx\sqrt{3\Delta}$. In
Figures~\ref{fig:sw}a and~b, we take $h=0.2$, corresponding to a
magnetic field of about $6\,\rm T$.

\begin{figure}
\vspace*{-\baselineskip}
\centerline{\includegraphics[width=7cm]{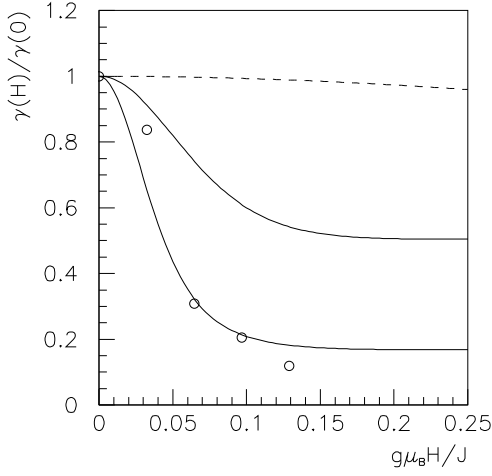}}
\caption{Magnetic-field dependence of the $\gamma$-coefficient. Bottom
	solid line: field applied perpendicular to the c-axis, top
	solid line: field parallel to the c-axis, both at $k_{\rm
	B}T=0.02\,J\approx0.5\,\mbox{K}$. Dashed line: field
	dependence at $k_{\rm B}T=0.2\,J\approx5\,\mbox{K}$. The open
	circles denote the data from
	Ref.~\protect\cite{helfrich95}. The ratio $\Delta$ of
	interchain- to intrachain-coupling is $10^{-4}$.}
\label{fig:goh} 
\end{figure}
The scenario outlined above allows for an at least qualitative
understanding of the behaviour of the specific heat $C_V$ in a
magnetic field as follows. For temperatures $k_{\rm
B}T/J\ll\sqrt{\Delta}$, the crystal behaves ``three-dimensional'',
i.e., $C_V\propto T^3$. If $k_{\rm B}T/J>\sqrt{\Delta}$, the
$T$-dependence becomes linear, $C_V\approx\gamma T$, because then
interchain-excitations are thermally populated, and only the
quasi-one-dimensional intrachain-excitations contribute. In
Figure~\ref{fig:goh} we show the field dependence of the
$\gamma$-coefficient at $k_{\rm B}T/J=0.02$ (solid lines) and $k_{\rm
B}T/J=0.2$ (dashed line). In a magnetic field parallel to the chains,
$\gamma$ is reduced by a factor of $2$. In this case, there is still
the above-mentioned zero-energy mode giving the main contribution to
$\gamma$. If the field direction points perpendicular to the chains,
$\gamma$ decreases rapidly as a function of $\mathbf h$ due to the
opening of the gap in $\omega_{Lc}(\mathbf q)$. This rapid decrease
compares quite well with the experiment---the experimental values are
drawn as open circles in the plot~\cite{helfrich95}. The experiment
was done using a multiple-domain crystal, so one would expect the data
to fall between the two extremes, $\mathbf h$ parallel and $\mathbf h$
perpendicular to the c-axis.

\begin{figure}
\vspace*{-\baselineskip}
\centerline{\includegraphics[width=7cm]{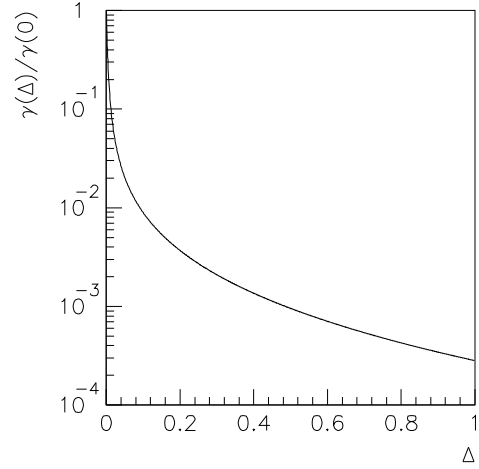}}
\caption{Dependence of the $\gamma$-coefficient at $k_{\rm
	B}T=0.02\,J\approx0.5\,\mbox{K}$ on the ratio $\Delta$ of
	interchain- to intrachain-coupling. Note the logarithmic scale on
	the ordinate.}
\label{fig:god}
\end{figure}
At temperatures $k_{\rm B}T/J\gg\sqrt{\Delta}$, the impact of $\mathbf
h$ on $\gamma$ is almost completely suppressed, see the dotted line in
Figure~\ref{fig:goh}. Compared with the experimental behaviour of
$\gamma(T)$ at sufficiently large fields, this provides us with an
order of magnitude for the interchain coupling, i.e., we have
$\Delta=\mathcal O(10^{-4})$ in order to describe the experiment.  The
strong dependence of the $\gamma$-coefficient on $\Delta$ is shown in
Figure~\ref{fig:god}. $\gamma$ decreases by almost four orders of
magnitude upon variation of $\Delta$ from zero to one, i.e., were
$\Delta>\mathcal O(10^{-2})$, the characteristic heavy-fermion
behaviour of \mbox{Yb$_\textrm{\scriptsize 4}$As$_\textrm{\scriptsize
3}$} would not be present.

We conclude that the spin-chain model for
Yb\textsubscript4As\textsubscript3 proposed in \cite{fulde95} and its
extension discussed here is in excellent aggreement with the neutron
scattering results of Kohgi et al.\ and the field dependence of the
$\gamma$-coefficient found by Helfrich et al.  In particular it
explains very well why this field dependence sets in only at low
temperatures. This is due to a new energy scale $\sqrt{JJ'}\ll J$ set
by the anisotropy gaps in the excitation spectrum which are caused by
the interchain coupling. From the behaviour of $\gamma(\mathbf H)$ one
would obtain $J'/J\simeq 10^{-4}$. In this case it would however be
impossible to directly observe the branches of excitation propagating
transversely to the chain direction (Fig.~\ref{fig:sw}) by neutron
scattering. We therefore suggest that magnetic resonance methods which
could determine the $\Gamma$-point excitation energies would be very
helpful in determining $J'$ and especially $d$ more
accurately. Although Yb\textsubscript4As\textsubscript3 is
semimetallic this should not be impossible since the residual
resistivity (about $1\,\mbox{mOhm-cm}$) is quite large in this
compound.

\section*{Acknowledgements}
We thank R.~Helfrich for allowing us to use his data prior to
publication and M.~Kohgi for interesting discussions and bringing his
unpublished data to our attention.


\begin{thebibliography}{99}
\bibitem{bonville94}{\sc Bonville P., Ochiai A., Suzuki T.\ {\rm and}
	Vincent E.},
	{\em J.~Phys.~I}, {\bf 4} (1994) 595.
\bibitem{helfrich95}{\sc Helfrich R., Steglich F.\ {\rm and} Ochiai A.},
	unpublished.
\bibitem{ochiai90}{\sc Ochiai A., Suzuki T.\ {\rm and} Kasuya T.}, 
	{\em J.~Phys.~Soc.~Jpn.}, {\bf 59} (1990) 4129.
\bibitem{goto95}{\sc Goto T.}, unpublished.
\bibitem{mori95}{\sc M\^ori N., Takahashi H., Okunuki Y., Nakai S.,
	Kashiwakura T., Kamata A., Teduka H., Kozuka Y., Yokohama Y.,
	Haga Y., Ochiai A., Suzuki T.\ {\rm and} Nomura M.},
	preprint.
\bibitem{krolas95}{\sc Rams M., Kr\'olas K., Tomala K., Ochiai A.\ 
	{\rm and} Suzuki T.},
	preprint.
\bibitem{kohgi95}{\sc Kohgi M.},
	unpublished.
\bibitem{suga95}{\sc Suga S.},
	unpublished.
\bibitem{reinders93}{\sc Reinders P.~H.~P., Ahlheim U., Fraas K.,
	Steglich F.\ {\rm and} Suzuki T.},
	{\em Physica B}, {\bf 186--188} (1993) 434.
\bibitem{fulde95}{\sc Fulde P., Schmidt B.\ {\rm and} Thalmeier P.},
	{\em Europhys.~Lett.}, {\bf 31} (1995) 323.
\bibitem{okeeffe77}{\sc O'Keeffe M.\ {\rm and} Andersson S.},
	{\em Acta Cryst.}, {\bf A33} (1977) 914.
\bibitem{cloizeaux62}{\sc des Cloizeaux J.\ {\rm and} Pearson J.J.},
        {\em Phys. Rev.}, {\bf 128} (1962) 2131.
\bibitem{affleck89}{\sc Affleck I., Gepner D., Schulz H.\ {\rm and}
	Ziman T.},
	{\em J.~Phys.~A}, {\bf 22} (1989) 511.
\bibitem{hallberg95}{\sc Hallberg K., Horsch P.\ {\rm and}
	Mart\'\i nez G.},
	{\em Phys.~Rev.~B}, {\bf 52} (1995) R719.
\bibitem{chubukov90}{\sc Abarzhi S.\ {\rm and} Chubukov A.},
	{\em J.~Phys.:~Condens.~Matter}, {\bf 2} (1990) 9221.
\bibitem{rastelli94}{\sc Rastelli E.\ {\rm and} Tassi A.},
	{\em Phys.~Rev.~B}, {\bf 50} (1994) 16475.
\bibitem{shiba93}{\sc Shiba H.\ {\rm and} Nikuni T.},
	{\em J.~Phys.~Soc.~Jpn.}, {\bf 62} (1993) 3268.
\bibitem{shiba94}{\sc Shiba H.\ {\rm and} Nikuni T.},
	preprint.
\end{thebibliography}
\end{document}